\documentclass[showpacs,preprint]{revtex4}
\usepackage{amsmath}
\usepackage{graphicx}
\usepackage{amsfonts, amssymb}
\usepackage{gensymb}
\usepackage{cool}

\usepackage{float}

\def\be{\begin{equation}}
\def\ee{\end{equation}}

\def\Hef{\vec{H}_{\mbox{{\tiny eff}}}}

\def\hef{\vec{h}_{\mbox{{\tiny eff}}}}

\def\kpar{k_u}

\def\vf{\varphi}
\usepackage{color}

\usepackage{ulem}

\begin{document}

\title{ On the speed of domain walls in thin nanotubes: the transition from the linear  to the magnonic regime}

\author{  M. C. Depassier}
\affiliation{Instituto de F\'\i sica, Pontificia Universidad Cat\'olica de Chile \\ Casilla 306, Santiago 22, Chile}



\begin{abstract}
Numerical simulations of domain wall propagation in thin nanotubes when an external magnetic field is applied along the nanotube axis have shown an unexpected behavior described as a transition from 
a linear to   a magnonic regime.  
As the applied magnetic field increases,  the initial 
regime of linear  growth of the speed with the field  is  followed by a sudden change in slope accompanied by the emission of spin waves.  In this work an analytical formula for the speed of the domain wall that explains this behavior is derived by means of an asymptotic study of the Landau Lifshitz Gilbert equation for thin nanotubes. We show that the dynamics can be reduced to  a one dimensional hyperbolic reaction diffusion equation, namely, the damped double Sine Gordon equation, which shows the  transition  to the magnonic regime as the domain wall speed approaches the speed of spin waves.  This equation has  been previously found to describe domain wall propagation in weak ferromagnets with the    mobility  proportional to the Dzyaloshinskii-Moriya interaction constant,   for Permalloy  nanotubes the mobility is  proportional to the nanotube radius. 
  \end{abstract}


\maketitle

\section{Introduction}

 Magnetic domain wall propagation is a subject of much current interest due to its possible applications in magnetic memory devices. Understanding and controlling the motion of  domain walls is essential for applications.  In the micromagnetic approach, the magnetization is governed by the Landau Lifshitz Gilbert (LLG) equation \cite{Landau1935,Gilbert1956}
 \be\label{LLG}
 \frac{\partial \vec m }{ \partial t} = - \gamma_0  \vec m \times \Hef(\vec m) + \alpha \vec m \times \frac{\partial \vec m }{ \partial t}
 \ee
 where $\vec m$ is the unit magnetization vector, that is,  the magnetization $\vec M = M_s \vec m$, where $M_s$ is the constant saturation magnetization, a property of the material.
 The constant  $\gamma_0=  | \gamma | \mu_0$,  where $\gamma$ is the gyromagnetic ratio of the electron and  $\mu_0$ is the magnetic permeability of 
vacuum. The parameter $\alpha>0$ is the dimensionless phenomenological Gilbert damping constant.  The effective magnetic field $\Hef$ includes the physical interactions and the external applied field $\vec H_a$.
The different physical phenomena that must be included in the effective field and the geometry of the ferromagnetic material together with the intrinsic nonlinearity of the problem imply that exact analytical solutions are generally nonexistent so that numerical and approximate analytic methods have been developed to understand experimental results and predict new phenomena. The exact solution of Walker \cite{Walker,ScWa1974} developed for an infinite medium with an easy axis, a local approximation for the demagnetizing field, including exchange interaction and under the action of an external magnetic field along the easy axis,  shows that when the applied field is small,  the speed of the domain wall increases linearly with the field. When the applied field reaches a critical value, the Walker field $H_{\text{w}},$  the magnetization enters into a precessing motion.  This behavior, which is  encountered even when additional physical effects and different geometries are studied, puts a limit to the maximum speed that a domain wall can achieve. 

For applications it is desirable to have stable domain walls and to reach high propagation velocities.  For such purpose different  physical effects and geometries have been considered. Numerical simulations for thin Permalloy nanotubes under the action of an external field along the nanotube axis showed unexpected behavior \cite{Yan2011,Hertel2016}.  For small fields the speed increases linearly with the field, reaching a plateau at relatively low applied field and very high velocity.  No instability nor Walker breakdown of  the domain wall was observed for this material in the parameter regime studied. This unexpected behavior occurs  for a specific chirality of the domain wall, namely  right handed domain walls,  
 for which the radial component of the magnetization remains small throughout the motion \cite{Hertel2016}.  

The main result of this manuscript is the derivation of an analytical expression for the speed of  the domain wall which explains the linear increase at small fields, the reaching of a plateau and the high values of the velocity. For Permalloy, a material of negligible uniaxial anisotropy,  we find
that the speed is given by 
\be\label{vper}
 v = \frac{ \gamma_0  R H_a}{\sqrt{ \alpha^2 + \mu_0 R^2 H_a^2/ (2 A)}},
 \ee
where $A$ is the exchange constant \cite{Hubert} and R the thin nanotube radius.
For small applied field  we recover the linear regime \cite{Goussev2014},
  \be\label{vw}
 v_{\text{L}} = \frac{\gamma_0}{\alpha} R H_a,
 \ee
 whereas for large applied field the speed tends to the constant value
 \be\label{vinfty}
 v_{\infty} = \gamma_0 \sqrt{\frac{2 A}{\mu_0}} = 1006 \, \text{ms}^{-1} \text{  for Permalloy}
 \ee
 which we identify with the minimal phase speed of spin waves. 
 Following the notation used  for weak ferromagnets, notice that Eqn.  (\ref{vper}) can be written as $ v= \mu  H_a/\sqrt{1 + \mu^2 H_a^2/v_{\infty}^2}$  with mobility $\mu =  (dv/dH_a )|_{H_a=0} = \gamma_0 R /\alpha$.
 The rise of the speed with the field is very fast; for a Permalloy nanotube of radius $R= 55$ nanometers, for an external field  field $B = \mu_0 H_a = 2$  mT,  Eqn. (\ref{vper})  yields $v = 893$ m s$^{-1}$. 
 
 The  suppression of the Walker breakdown together with a slowdown of a domain wall as it approaches the phase speed of spin waves  has been encountered previously in different problems.  In antiferromagnets with Dzyaloshinskii-Moriya interaction (DMI)   the  mobility was found to be proportional to the  DMI constant \cite{Gyorgy1968,Zvezdin1979,Papanicolaou1997}. See the recent review  \cite{Galkina2018} for additional references.   
 Similar behavior was found  in rough nanowires \cite{Nakatani2003,Thiaville2006}  and  in nanowires  with a  strong hard axis perpendicular to the wire when  an external field is  applied along the wire \cite{Wieser2004,   Wieser2010, Wang2014}.  A theoretical explanation for the effect of spin waves on  Bloch walls was given in \cite{Bouzidi} where it was shown that the transition to the magnonic regime may occur before or after the Walker breakdown depending on the parameters of the problem. See also \cite{Akhiezer1967, Baryakhtar1985}.
  In all these works bulk matter or thin films were the subject of study. For Permalloy  nanotubes the numerical simulations  of \cite{Yan2011,Hertel2016} show  that the sudden change of slope in the rate of increase of the speed of domain walls is accompanied by Cherenkov spin wave emission once the DW speed exceeds the phase speed of the spin waves.
 
   In the present work we study theoretically the DW propagation in Permalloy nanotubes and  find that curvature acts as an additional anisotropy and  plays an  equivalent role to the  DMI in  weak ferromagnets. The parallel between curvature of a nanotube and  DMI has been observed in \cite{Goussev2016,Gaididei2014,Ot2016} among others.  In \cite{Ot2016} it is shown that  the analytical expression of the dispersion relation for spin waves in a nanotube  has the same mathematical form as the dispersion relation for spin waves  in thin films with DMI.  Here we find this mathematical analogy  in the  mobility of the domain wall. 
   See \cite{Stano2018} for a recent  comprehensive review on the dynamics of magnetic nanotubes.

    Although simulations have been carried out for Permalloy,  in the derivation below we will allow  a material with non negligible uniaxial anisotropy for  greater generality.

 \section{Statement of the problem}
 
\begin{figure}[h]\label{fig:coord}
\centering
\includegraphics[width=0.35 \textwidth]{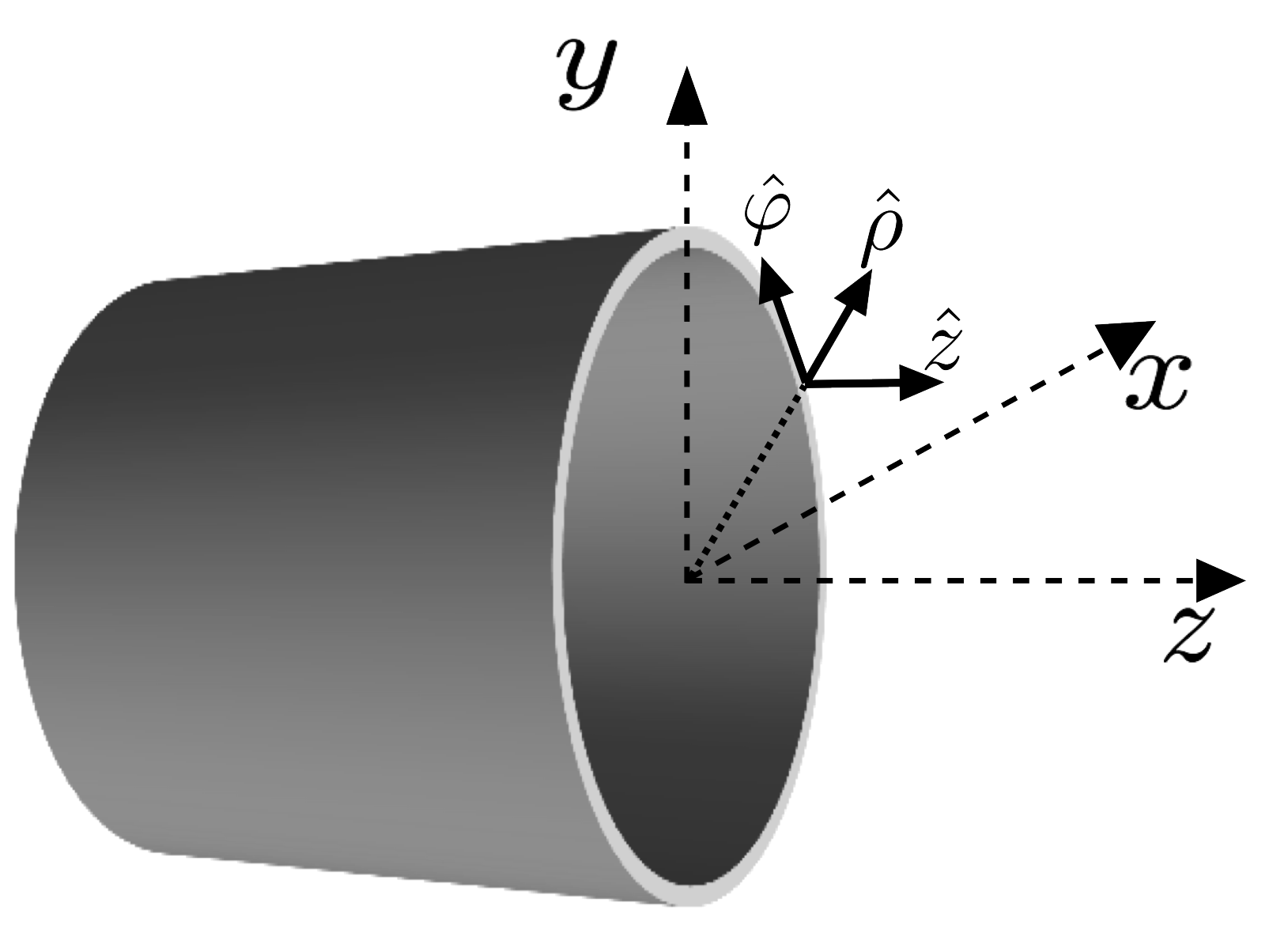}
\caption{Cylindrical coordinate system in the nanotube.}
\end{figure}

Consider  a thin nanotube with an easy direction along the nanotube axis  which we choose as the $z$ axis.       
The dynamic evolution of the magnetization  is governed  by the LLG equation (\ref{LLG}).
A  right handed  orthogonal cylindrical coordinate system $( \rho, \varphi, z)$   is introduced as shown in Fig.\ref{fig:coord} in terms of which the   unit magnetization vector is written as   $\vec m = m_{\rho}(\rho,\varphi,z)  \hat\rho + m_{\varphi}(\rho,\varphi,z) \hat \varphi + m_z(\rho,\varphi, z) \hat z$.

For sufficiently thin tubes the demagnetizing field can be approximated by a local expression  with the saturation magnetization acting as an effective radial hard axis anisotropy \cite{Carbou2001, Kohn2005, Goussev2014}. 
In this approximation and including  exchange energy, uniaxial anisotropy energy, demagnetization energy and Zeeman energy, 
the micromagnetic energy can be written as  \cite{Hubert,Goussev2014}
 \begin{equation}
 E=\int_{\Omega}d^3x ( A |\nabla \vec m|^2 + K_u(1 - m_z^2) + \frac{\mu_0 M_s^2}{2} m_{\rho}^2 - H_a m_z),
 \end{equation}
 where $\Omega$ is the material  volume of the nanotube,    $A$ is the exchange constant,   $K_u$ the uniaxial anisotropy and an external field $\vec H_a =H_a \hat z$ has been applied along the axis. The effective field is given by
   $$
 \Hef = - \frac{1}{\mu_0 M_s} \frac{\delta E }{\delta \vec m}.
 $$
 In a very thin  nanotube  variations of the magnetization with radius may be neglected  so that the unit magnetization depends only on the polar coordinate $\varphi$ and the axial position  $z$.  With $\vec m = \vec m(\varphi,z)$,   the  effective magnetic field can be written  as \cite{Goussev2014}
\be\label{H1}
\vec H_e = \frac{2 A}{\mu_0 M_s} \left[ \frac{1}{R^2} \frac{\partial^2  \vec m}{\partial \varphi^2} +  \frac{\partial^2  \vec m}{\partial z ^2} \right] + \frac{2 K_u}{\mu_0 M_s} m_z \hat z - M_s m_{\rho} \hat \rho  + H_a \hat z.
\ee
Introducing $M_s$ as unit of magnetic field, and introducing the dimensionless space  and time variables   $\xi =  z/R$  and $\tau= \gamma_0  M_s t$  we rewrite equations (\ref{LLG}) and (\ref{H1}) in dimensionless form as
\be\label{LLG1}
 \frac{d \vec m }{ d \tau} = -   \vec m \times \hef + \alpha \vec m  \times \frac{d \vec m }{ d\tau }
 \ee
 with
\be \label{hef}
\hef=  A_0  \left[ \frac{\partial^2 \vec m}{\partial \varphi^2} +  \frac{\partial^2  \vec m}{\partial \xi ^2} \right]  + \kpar m_z \hat z - m_\rho \hat\rho + h_a \hat z 
\ee
where $h_a$ is the dimensionless applied field.  The dimensionless numbers that have appeared are $\kpar= 2 K_u/(\mu_0 M_s^2)$ and $A_0$,   the square of the  ratio  between the  exchange length 
$l_{ex} = \sqrt{2 A/\mu_0 M_s^2}$  and the radius,  that is,  $A_0 = 2 A/ (\mu_0 M_s^2 R^2)$.  Equations (\ref{LLG1})  and (\ref{hef})  describe the dynamics of the problem.

Numerical simulations \cite{Yan2011,Hertel2016} have been performed for Permalloy for which the exchange constant $A= 1.3 \times 10^{-11}$ J m$^{-1}$, $M_s =  8 \times 10^5 $A m$^{-1}$, $K_u \approx 0$ and the external applied field does not exceed $ 10^{-2} M_s$.  The nanotube used in simulations has inner radius $R$,  and width $w$ with $w<<R.$  Here we neglect the variations with radius and consider the range  $R = 55-100 \times 10^{-9}$ m.  The vacuum permeability $\mu_0 = 4 \pi \times 10^{-7} $N A$^{-2}$ so that $\mu_0 M_s \approx 1 $T. We take the value $\gamma_0 = 2.21 \times 10^5$ s$^{-1}$T$^{-1}$.   For Permalloy the exchange length is $l_{ex} = 5.68$ nm  and for a radius of $80$ nm $l_{ex}/R = 0.071$. 
The uniaxial anisotropy vanishes, $\kpar=0$,  and  the dimensionless applied field  is  in the range $0< h_a < 10^{-2}$.  The damping parameter $\alpha$ in numerical simulations ranges from $0.01$ to $0.03$.

 In the numerical simulations of Permalloy nanotubes \cite{Yan2011,Hertel2016} it is observed that a right handed vortex-type domain wall is stable against the Walker breakdown when driven with a magnetic field. The handedness of the domain wall in \cite{Yan2011, Hertel2016} is defined in relation to the applied field,  a propagating domain wall is called right handed if $(\vec m \times \vec H_a)\cdot \hat\rho >0$. Left handed domain walls become unstable and  convert into the other, stable chirality.   In this work we are interested in the speed of the stable DW, which will be selected through the scaling in the asymptotic solution.
 
 \section{Asymptotic solution}
 
 In this section we perform an asymptotic analysis of the LLG equation to find a reduced model for the evolution of the domain wall as the applied field increases. The reduced model will be valid for a restricted parameter range which is chosen based on the numerical results described above for Permalloy.  We are interested in right handed vortex walls for which the radial component of the magnetization is small, $m_{\rho} \ll 1$ \cite{Yan2011, Hertel2016}.   Introducing a small dimensionless parameter $\epsilon$ we write this condition as
 \be
  m_{\rho}= \epsilon \tilde m_{\rho}.
  \ee
The normalization condition $\vec m^2 = 1$ becomes
  \be \label{m2A}
m_{\vf}^2 + m_z^2 = 1 - \epsilon^2  {\tilde m_{\rho}}^2.
  \ee
   We will model a situation in which the ratio  $l_{ex} /R$ and the Gilbert constant are of the same order in $\epsilon$  as the radial component of the magnetization. We assume that   the applied field and uniaxial anisotropy  are of an order smaller.  Let then
     \be
  A_0  = \epsilon^2 \tilde A, \qquad  \kpar = \epsilon^2 \tilde \kpar, \qquad h_a = \epsilon^2 \tilde h_a, \qquad \alpha= \epsilon \tilde \alpha.
\ee
  It is found that a consistent  asymptotic approach can be obtained if a new time scale $s = \epsilon \tau$ is introduced. With these scalings,  the components of the effective magnetic field can be written as 
   
   \begin{subequations}\label{field}
 \begin{align}
\begin{split}
 (\hef)_{\rho} 
  &=  - \epsilon   \tilde m_{\rho} 
  - 2 \epsilon^2   \tilde A \frac{\partial m_{\vf} }{\partial \varphi} +
   \epsilon^3 \tilde A  \left(  
    \nabla_s^2 \tilde m_{\rho}  -\tilde m_{\rho}\right)         \\
      &=  \epsilon  H^0_{\rho}   +  \epsilon^2  H^1_{\rho} + \epsilon^3  H^2_{\rho} ,  
\label{hr} 
 \end{split} \\
 \begin{split}
 (\hef)_{\varphi}  &= \epsilon^2  \tilde A  (  \nabla_s^2 m_{\varphi} -   m_{\varphi} )  + 2 \epsilon^3 \tilde A   \frac{\partial \tilde m_{\rho} }{\partial \varphi}  \\
  &=  \epsilon^2 H^1_{\varphi} +   \epsilon^3 H^2_{\varphi}, 
  \end{split}
  \label{hfi} \\
 (\hef)_{z}  &= \epsilon^2 \left( \tilde h_a + \tilde A  \nabla_s^2 m_{z}   +   \tilde \kpar m_z  \right) =  \epsilon^2  H^1_{z}, 
 \label{hz}
 \end{align}
  \end{subequations}
where $\nabla_s^2 = \partial_{\xi \xi} + \partial_{\varphi \varphi}$  and where we grouped terms according to the power of $\epsilon$ so that  $ H^0_{\rho}  = -  \tilde m_{\rho} $, $H^1_{\varphi}= \tilde A (\nabla_s^2 m_{\varphi} - m_{\varphi})$ and 
  $H^1_{z} = \tilde h_a + \tilde A  \nabla_s^2 m_z +  \tilde \kpar m_z .$  In obtaining these expressions for the effective field  the property $\partial\hat \rho/\partial \varphi = \hat \varphi, \, \partial\hat \varphi /\partial \varphi = - \hat \rho$ is used.

 Introducing the scaling  for $\alpha$ and  $m_{\rho}$ in the LLG equation, we obtain at leading order in $\epsilon,$
  \begin{subequations}
  \label{eqs}
 \begin{align}\label{mro}
 \dot{\tilde m}_{\rho} &= -  ( m_{\varphi} H_z^1 - m_z H_{\varphi}^1 ) +\tilde \alpha (  m_{\varphi} \dot{m}_z - m_z \dot{m}_{\varphi} ),
 \end{align}
 \begin{align}\label{mfi}
  \dot{m}_{\varphi}  &= - m_z  H_{\rho}^0,  
   \end{align}
 \begin{align}\label{mz}
 \dot{m}_{z} &=   m_{\varphi}  H_{\rho}^0,     \end{align} 
 \end{subequations}
where a dot represents a derivative with respect to the scaled time variable $s$ and the subindices represent the components of each vector.

The normalization condition (\ref{m2A})  implies that, at leading order,  we may write
\be
m_{\varphi} = \sin\theta(\xi,\varphi,s), \qquad \qquad m_z = \cos\theta(\xi ,\varphi,s).
\ee
It follows then that equations (\ref{mfi}) and (\ref{mz}) are equivalent and imply
\be\label{uno}
\dot\theta = -  H_{\rho}^0 = \tilde m_{\rho}.
\ee

Replacing the value of $\tilde m_{\rho}$ from (\ref{uno}) in (\ref{mro}) together with the expressions for the effective field $H_{\varphi}^1, H_z^1$,  the evolution equation for $\theta$ is found to be 
\be\label{HRDE}
 \ddot \theta + \tilde \alpha \,\dot\theta =  \tilde A \left( \theta_{\xi\xi} +  \theta_{\varphi \varphi} \right)  
- \sin\theta \left( \tilde h_a +  ( \tilde A + \tilde \kpar ) \cos\theta \right),
\ee
where the subscripts in $\theta$  denote derivatives with respect to $\xi$ and $\varphi$ respectively.   Notice that one may go back to the original unscaled variables and the small parameter $\epsilon$ cancels out. 

In what follows we study  cylindrically symmetric domain walls, for which $\theta_{\varphi}=0$   and identify the evolution equation with the  damped double Sine Gordon equation, 
\be\label{final}
 \frac{\partial^2  \theta}{\partial \tau^2}  +  \alpha \frac{\partial \theta}{\partial \tau}  =   A_0  \theta_{\xi\xi}  
- \sin\theta \left(  h_a +  ( A_0 + \kpar ) \cos\theta \right), 
\ee
a particular case of hyperbolic reaction diffusion equation,  for which the existence and stability of traveling waves  have been studied  rigorously in \cite{Hadeler1988,Gallay2009}.

  This equation has been derived in the analysis of domain wall propagation  in  weak ferromagnets, \cite{ Gyorgy1968,Zvezdin1979,Mikeska1980,Papanicolaou1997} and in systems with a strong easy plane \cite{How1989, Wieser2010}.   In \cite{Papanicolaou1997} the dependence of mobility on the Dzyaloshinskii constant is derived with great detail.  A common  feature in these problems is the sudden  decrease in the rate of increase of the speed with the applied field.

This equation has the same traveling wave solutions  as the reaction diffusion equation $\alpha\dot\theta = A_0  \theta_{\xi\xi}  
- \sin\theta \left(  h_a +  ( A_0 + \kpar ) \cos\theta \right)$ but with velocity $c = c_{\text{r}} / \sqrt{1 + c_{\text{r}}^2/A_0}$ where $c_{\text{r}}$ is the speed of fronts of  the reaction diffusion equation \cite{Hadeler1988}.   We give the explicit expression for the head to head (HH) domain wall, the tail to tail solution is similar. The HH solution is found to be  the usual domain wall profile, 
 \be\label{HH}
  \theta(\xi,t) = 2 \arctan\left[ \exp \left( \frac{\xi - c \tau}{\Delta} \right) \right]
 \ee
 with the speed $c$ and domain wall width $\Delta$  given by  \be\label{speedad}
 c = \sqrt{ \frac{A_0}{A_0+ \kpar}}\,\, \frac{h_a}{ \sqrt{\alpha^2 +    (A_0+\kpar)^{-1}  \, h_a^2}},  \qquad \Delta = \frac{\alpha c}{h_a}.
 \ee 
The leading order magnetization $\vec m = m_{\varphi} \hat\varphi  + m_z \hat z$ is given by
\be
\vec m = \sech \left( \frac{\xi - c \tau}{\Delta} \right) \hat \varphi - \tanh\left ( \frac{\xi - c \tau}{\Delta} \right) \hat z.
\ee
The external field  is applied along the $z$ axis so the magnetization is  a right handed ($m_{\varphi}\ge 0$)  head to head domain wall as defined in \cite{Hertel2016}. The small radial component of the magnetization is calculated from \eqref{uno}.

For small applied field we recover the linear regime, that is, the speed increases linearly with the field, and the domain wall width tends to a constant value, that is, 
\begin{equation}
\lim_{h_a\rightarrow 0} \, c = \sqrt{ \frac{A_0}{A_0+ \kpar}}\,\, \frac{h_a}{ \alpha},   \quad \lim_{h_a\rightarrow 0} \, \Delta = \sqrt{\frac{A_0}{A_0 + \kpar}}.
\end{equation}
In this limit the dynamics is primarily governed by the reaction diffusion equation $\alpha \theta_{\tau}   =   A_0  \theta_{\xi\xi}  
- \sin\theta \left(  h_a +  ( A_0 + \kpar ) \cos\theta \right)$ as already found in \cite{Goussev2014}.

In terms of the physical parameters the dimensional  domain wall width  for small field  $\delta =  \Delta R$ and speed $v_L$ can be written as
$$
\delta =  \sqrt{\frac{A}{\frac{A}{R^2} + K_u}}, \qquad v_L = \gamma_0 \frac{H_a}{\alpha} \delta
$$
For Permalloy, $K_u=0$ and $\delta = R$ in agreement with the results for a static domain wall in a  thin nanotube \cite{La2010}.  The speed $v_L$ coincides with the low field Walker solution $v_W= \gamma_0 H_a \sqrt{A/K}/\alpha$, with $K$ an effective anisotropy $A/R^2$.
In the limit of large radius the domain wall width for an in plane magnetized  thin film, $\sqrt{A/K_u} $ is recovered.

For large applied field the speed tends to a constant value and the domain wall width decreases as the field increases, 
\be
\lim_{h_a\rightarrow \infty} \, c = \sqrt{ A_0}     \quad \lim_{h_a\rightarrow \infty} \, \Delta = \frac{\alpha  \sqrt{A_0}}{h_a}.
\ee
 This limiting value for the speed corresponds to the minimal value of the phase velocity for spin waves, $v_{p_ \text{\tiny{min}}}$. In effect, consider a material like Permalloy with vanishing uniaxial anisotropy for simplicity. The dispersion relation for the DSG equation (\ref{final}), for vanishing damping and vanishing applied field, is given by $\omega_{\text{\tiny DSG}}=  \sqrt{A_0} \sqrt{1 + k^2}$, so that the phase speed is a decreasing function of $k$ which tends asymptotically to 
 $\sqrt{A_0}$ as $k$ grows. The full dispersion relation for    spin waves in a thin nanotube, in the absence of damping and applied field, with vanishing  uniaxial anisotropy,  is given by \cite{GoLaNu2010} 
\be
 \omega =\frac{\sqrt{A_0}}{2} \sqrt{(1 + k^2)  + A_0 (1 + k^2)^2 }
 \ee
 in the units  used in this work. We see that for small $A_0$  the full dispersion relation coincides,  up to a constant,  with $\omega_{\text{\tiny DSG}}$.

The evolution equation (\ref{final}) shows the transition from the low field regime where the speed of the domain wall  increases linearly with the field to the regime where the domain wall speed approaches 
 $v_{p_ \text{\tiny{min}}}$  and is slowed down by emitting spin waves.  
 In order to capture the large applied field regime where the DW speed exceeds $v_{p_ \text{\tiny{min}}}$  and Cherenkov emission occurs a different scaling is needed. The DW width shrinks with increasing field, $\Delta \approx \alpha \sqrt{A_0}/h_a$, which indicates that at larger fields   a new scaling for the longitudinal coordinate $\xi$ is required.  The damped DSG equation  captures  the  emission of spin waves that occurs below but close to  $v_{p_ \text{\tiny{min}}}$.

A different transition occurs at $h_a^{\text{\tiny{KPP} }}= 2 A_0$ when the speed of the reaction diffusion equation $c_r$ changes from a pushed to a pulled or KPP front \cite{DepassierEPL} and $c_r$ becomes proportional to the square root of the applied field.

\begin{figure}
\centering
\includegraphics[width=0.45 \textwidth]{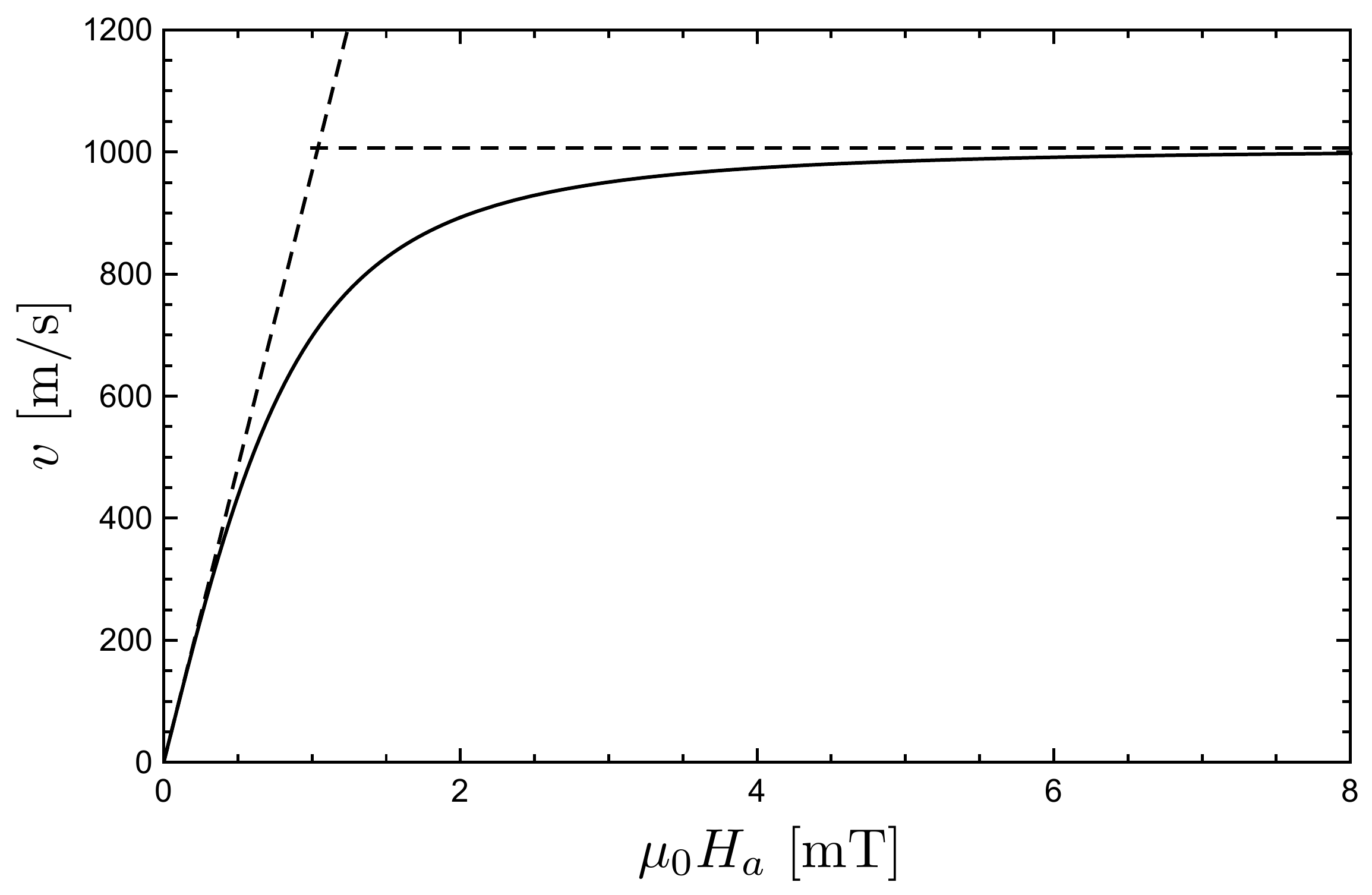}
\caption{Speed of the domain wall versus applied field in millitesla for a thin Permalloy nanotube of radius $R =  55 $ nanometers. At low field the speed increases linearly with the field, after a sudden change in slope the speed tends to a constant value $v_{\infty}$ at large fields. The dashed lines show the limiting speeds $v_{\text{L}}$ and $v_{\infty}$.}
 \label{fun}
\end{figure}

 In what follows consider Permalloy for which $\kpar =0$.   Going back to dimensional quantities, the speed of the domain wall for Permalloy is given by Eq. (\ref{vper}) with the limiting values at low and high fields Eq. (\ref{vw}) and Eq. (\ref{vinfty}).  In Fig. \ref{fun} the graph of the speed as a function of  the applied field  shows the gradual change from the linear to the magnonic regime. We have used the values given above for Permalloy.  
 
 An approximate  estimate of the field $H_a^*$ at which this transition occurs is obtained by the intersection $v_{\text{L}}(H_a^*) = v_{\infty}$  which yields
 $$
 H_a^* = \frac{\alpha}{R} \sqrt{\frac{ 2 A}{\mu_0}}.
 $$
For Permalloy we obtain $v_{\infty} = 1006$ m s$^{-1}$, and $B_a^* = \mu_0 H_a^* = 0.001 $ T.  The transition to the KPP regime occurs at a much higher field, $B_a^{\text{\tiny{KPP} }} = \mu_0  H_a^{\text{\tiny{KPP} }} =  0.021$ T and is not associated to the transition from the linear to the magnonic regime.  In this simple model the   order of magnitude of the speed and the value of the field at which the transition from the linear to the magnonic regime occurs   agrees with the order of magnitude of the numerical simulations of the LLG equation.

\section{Summary}

We studied the dynamics of a vortex domain wall  in a thin nanotube  by means of an asymptotic study of the Landau-Lifshitz Gilbert equation in a parameter regime based on  existing numerical simulations \cite{Yan2011,Hertel2016}.  
The numerical simulations on Permalloy nanotubes in a certain range of radii showed that when an external field is applied along the axis, domain walls of one type of chirality, for which the radial magnetization remains small during the motion, are stable and can reach high speeds. Initially the speed increases linearly with the applied field, and at higher fields the rate of increase is slowed down by the emission of spin waves. No Walker breakdown was observed in the parameter range considered in the numerical studies.  Domain walls of the opposite chirality are unstable and as the field increases they convert into DW of stable chirality.

The purpose of this work was  to understand analytically the behavior of the speed of domain walls of stable chirality as a function of the applied field.  Through  an asymptotic analysis  the LLG was reduced to the damped double sine-Gordon  equation from which an explicit analytic formula for the speed as a function of the applied field was obtained together with the leading order DW profile. This model captures the initial regime of linear growth of the speed followed by a slowdown in the rate of increase through the emission of spin waves before reaching the minimal phase speed of the spin waves, which is an upper bound on the speed of the DW in this model. The order of magnitude of the speed and the value of the applied field where the transition  from the linear to the magnonic regime occurs is in agreement with the numerical results of  \cite{Yan2011,Hertel2016}.  In order to reach higher fields and capture the Cherenkov spin wave emission process a different asymptotic regime is necessary. 

For Permalloy, which has vanishing uniaxial anisotropy, 
 the ratio of the exchange constant with the square of the radius of the nanotube $A/R^2$ plays the role of an effective uniaxial anisotropy  which leads to a mobility proportional to the nanotube radius.  In constrast, for weak ferromagnets the mobility is proportional to de Dzyaloshinkii-Moriya constant.  That the effect of curvature acts as an equivalent effective anisotropy  was already shown  in \cite{Goussev2014,Gaididei2014}, and  an analogy between the effect of DMI and curvature was found  in the dispersion relation of spin waves in a nanotube \cite{Ot2016}. The results in this manuscript show a similar effect  when studying the transition from the linear  to the magnonic DW  regime in nanotubes.  
 
 An analytical approach to the regime of higher field, where Cherenkov emission occurs, will be the subject of future study.

   \section{Acknowledgments}

This work was partially supported by Fondecyt (Chile) project 116--0856.

    \end{document}